\documentclass[reprint,amsmath,amssymb,aps,floatfix]{revtex4-2}
\usepackage{graphicx}
\usepackage{dcolumn}
\usepackage{bm}
\usepackage{miller}
\usepackage{natbib}
\usepackage{upgreek}
\usepackage{graphicx}
\usepackage{amsmath}
\usepackage{color}

\begin{document}

\preprint{APS/123-QED}

\title{Towards hexagonal $C_{2v}$ systems with anisotropic DMI:\\ Characterization of epitaxial Co\hkl(10-10)/Pt\hkl(110) multilayer films}

\author{Yukun Liu}
\author{Michael D. Kitcher}
\author{Marc De Graef}
\author{Vincent Sokalski}
\affiliation{
 Department of Materials Science \& Engineering, Carnegie Mellon University, Pittsburgh, PA, USA 15213
}

\date{\today}

\begin{abstract}
 
Ferromagnet/heavy metal (FM/HM) multilayer thin films with $C_{2v}$ symmetry have the potential to host antiskyrmions and other chiral spin textures via an anisotropic Dzyaloshinkii-Moriya interaction (DMI).  Here, we present a candidate material system that also has a strong uniaxial magnetocrystalline anisotropy aligned in the plane of the film. This system is based on a new Co/Pt epitaxial relationship, which is the central focus of this work: hexagonal closed-packed Co\hkl(10-10)\hkl[0001] $\parallel$ face-centered cubic Pt\hkl(110)\hkl[001]. We characterized the crystal structure and magnetic properties of our films using X-ray diffraction techniques and magnetometry respectively, including q-scans to determine stacking fault densities and their correlation with the measured magnetocrystalline anisotropy constant and thickness of Co.  In future ultrathin multilayer films, we expect this epitaxial relationship to further enable an anisotropic DMI while supporting interfacial perpendicular magnetic anisotropy. The anticipated confluence of these properties, along with the tunability of multilayer films, make this material system a promising testbed for unveiling new spin configurations in FM/HM films.

\end{abstract}

\maketitle

\section{\label{sec:intro}Introduction}

Magnetic skyrmions are promising candidates for future spintronic data storage and logic devices because of their small size, heightened stability, and high mobility \cite{Racetrack, Nagaosa, Perspective, Reservoir, Roadmap}. These structures are typically stabilized via the Dzyaloshinskii-Moriya  interaction (DMI) which emerges in magnetic systems with high spin-orbit coupling and no inversion symmetry \cite{Dzyaloshinskii, Moriya, Spontaneous}. In polycrystalline ferromagnet/heavy metal (FM/HM) thin films, only Néel skyrmions have been stabilized, due to the high symmetry of these systems \cite{Muhlbauer, RealSpace, FeIrSTM, Nagaosa, Perspective, Roadmap, Moriya, Gungordu, Hoffmann, Raeliarijaona}. However, recent research has shown that low-symmetry FM/HM thin film interfaces exhibit anisotropic DMI which, in turn, could stabilize antiskyrmions—the topological opposites of skyrmions \cite{Gungordu, Raeliarijaona, Camosib, Hoffmann, Huang}. Specifically, antiskyrmions are preferred over Néel skyrmions when the DMI has opposing signs along orthogonal crystallographic directions. To date, research on anisotropic DMI in FM/HM systems has focused on films based on a body-centered cubic (BCC) heavy metal 
\cite{Camosib, Hoffmann, Huang, CamosiMicroMag, CamosiHans, TZ}. Moreover, across the limited experimental studies carried out thus far, a suitable antiskyrmion host has not been found.

To expand this search, we present a system that exhibits the growth of hexagonal close-packed (HCP) Co on a face-centered cubic (FCC) Pt underlayer, resulting in a singular magnetocrystalline easy axis that lies in the plane of the film. When coupled with an (an)isotropic DMI and/or an interface-induced perpendicular easy axis, such in-plane uniaxial anisotropy is likely to broaden the range of spin configurations that can exist in FM/HM films.  In this work, we characterized the epitaxial relationship that enables such a system; investigated the prevalence of structural defects—namely, intrinsic stacking faults (SFs)—during the growth of films; and measured the magnetocrystalline anisotropy (MCA) by probing the resulting in-plane hysteresis.

\begin{figure}[!htbp]
\includegraphics[width=0.5\textwidth]{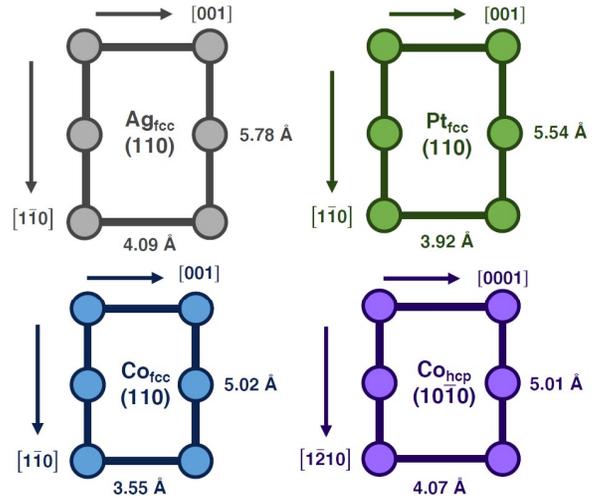}
\caption{\label{tab:LatticeMatching}
Orientation and interatomic relationship between  FCC Ag\hkl(110), FCC Pt\hkl(110), FCC Co\hkl(110), and HCP Co\hkl(10-10).}
\end{figure}

\section{\label{sec:methods}Methods}
All films were grown via DC magnetron sputtering in an Ar atmosphere at a fixed working pressure of 2.5 mTorr. Prior to deposition, single-crystal Si\hkl(110) substrates were HF-etched for 1 minute to remove any native oxide that would otherwise impede epitaxial growth.  Following Yang et al., we deposit a Ag buffer layer which is known to grow with the FCC \hkl(110) texture on Si\hkl(110) \cite{Laughlin}. This growth seed facilitates the deposition of our films at room temperature, thus avoiding the severe interlayer diffusion that occurs at elevated temperatures \cite{Laughlin}. Moreover, the degree of lattice mismatch between Ag and Pt is rather moderate ($\sim$ 4.3\%). Pt[10 nm]/Co[20 nm, 50 nm]/Pt[3 nm] thin films were subsequently deposited.
To identify the phases and textures of the different layers, $2\theta$-$\omega$ coupled scans and azimuthal ($\varphi$) scans were performed on the deposited thin films using a Phillips X'Pert Pro X-ray diffractometer with unfiltered Cu $K_\alpha$ radiation. Furthermore, the density of SFs in HCP Co for each sample was quantitatively determined from X-ray reciprocal space q-scans. Using measurements of the full width at half maximum ($w_h$) of different Co peak profiles along different directions, we extracted the broadening effects caused by different intrinsic SFs. The corresponding SF densities were subsequently calculated.

The magnetic anisotropy of the deposited thin films was examined using a Princeton Measurements alternating gradient force magnetometer (AGFM). $M$\textendash $H$ hysteresis curves were obtained for orthogonal in-plane directions and their corresponding magnetic anisotropy energy densities were calculated from the area between one half of the idealized hard axis loop and the $M$-axis.

\section{\label{sec:r&d}Results and Discussion}
\begin{figure}[!htbp]
\includegraphics[width=0.5\textwidth]{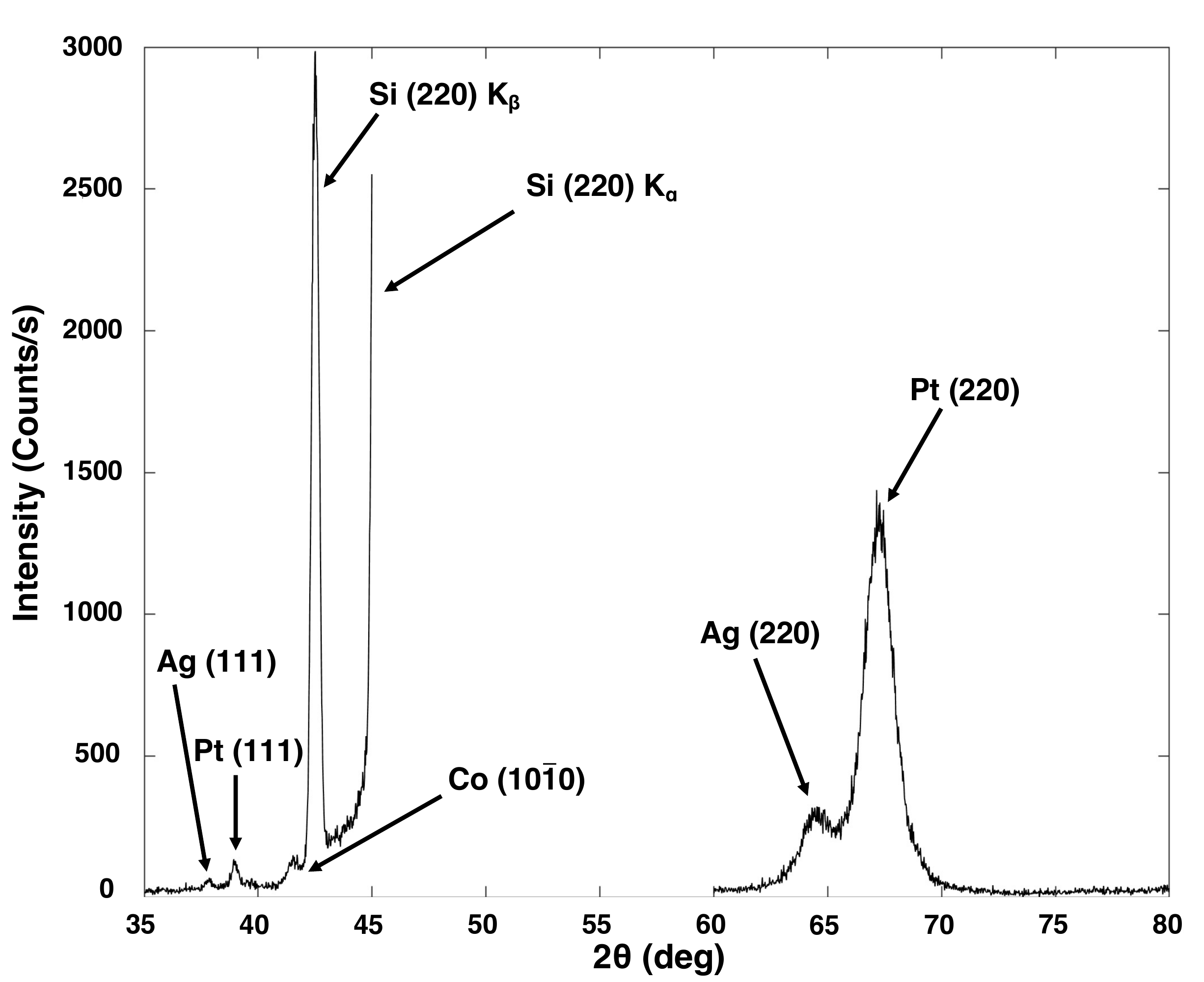}
\caption{\label{tab:2ThetaOmega}
Diffraction profiles of $2\theta$-$\omega$ coupled scans performed on the Co(20 nm)/Pt film.}
\end{figure}

\begin{figure}[!htbp]
\includegraphics[width=0.5\textwidth]{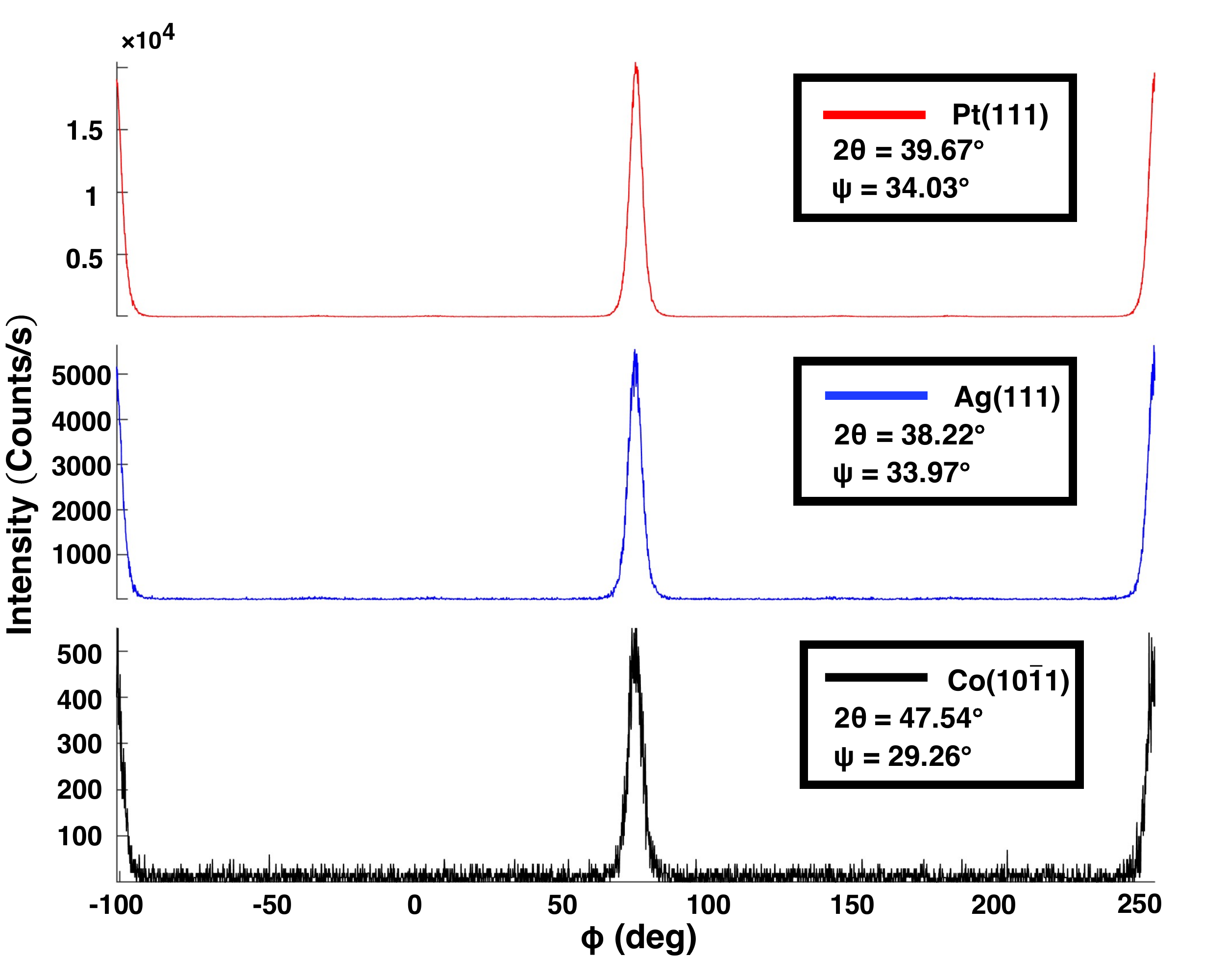}
\caption{\label{tab:Phicombined}
Diffraction profiles of azimuthal ($\varphi$) scans performed on the Co(20 nm)/Pt film.}
\end{figure}

The representative $2\theta$-$\omega$ coupled scan profiles presented in Figure \ref{tab:2ThetaOmega} show dominant HCP Co\hkl(10-10), FCC Pt\hkl(220), and FCC Ag\hkl(220) diffraction peaks, indicating that while Pt grows with the same preferred FCC \hkl(110) orientation of the Ag underlayer, Co adopts the HCP \hkl(10-10) texture. Here, we use Bravais-Miller indexing notation ($HKIL$) to represent hexagonal structures where $I = -(H + K)$. These results are corroborated by the azimuthal scan profiles in Figure \ref{tab:Phicombined} which reveal HCP Co\hkl(10-11) planes at the same azimuth as FCC \hkl(111) planes in Pt and Ag. Correspondingly, the epitaxial relationship between layers is Co\hkl(10-10)\hkl[0001] $\parallel$ Pt\hkl(110)\hkl[001] $\parallel$ Ag\hkl(110)\hkl[001] $\parallel$ Si\hkl(110)\hkl[001]. The strong \hkl(110) texture of the Ag growth seed in these films agrees with the work of Yang et al., who noted that a $4 \times 4$ mesh of Ag unit cells on a $3 \times 3$ mesh of Si unit cells has a lattice mismatch of less than 0.45\% \cite{Laughlin}. 

Given the prevalence of ``all-FCC" Co/Pt multilayers in magnetics research, it may seem counterintuitive that a cubic Pt\hkl(110) underlayer would give rise to a hexagonal Co layer. However, as shown in Figure \ref{tab:LatticeMatching}, the HCP Co\hkl(10-10) surface has significantly lower lattice mismatch with the Pt(110) surface along [001] as compared to the FCC Co equivalent, while the mismatches along the \hkl[1-10] of Pt are nearly identical.  Together with cobalt's thermodynamic preference for the HCP phase at room temperature, the potential for a lower strain drives the formation of the observed texture. However, the theoretical lattice mismatch along Pt \hkl[1-10] at the observed Co/Pt interface is considerable. We therefore explored the prevalence of intrinsic SFs in these films and their impact on the magnetocrystalline uniaxial anisotropy along the c-axis of Co. 

It has been established that intrinsic SFs in HCP crystals broaden diffraction peaks along \hkl[0001] when $H - K \neq 3N$, where $N \in \mathbb{Z}$. Furthermore, the degree of broadening depends on the parity of $L$. In particular, the normalized broadening, $\Delta w_{h, sf}$, of these peaks in reciprocal space and the density of SFs are related by:  
\begin{subequations}
\begin{align}
\label{eqn:equation 1a}
\Delta w_{h, sf} &= \frac{3\alpha+3\beta}{\pi}, \;L \;\mathrm{mod}\; 2 = 0;\\
\label{eqn:equation 1b}
\Delta w_{h, sf} &= \frac{3\alpha+\beta}{\pi}, \;L \;\mathrm{mod}\; 2 = 1,
\end{align}
\end{subequations}
where $\alpha$ and $\beta$ represent the densities of deformation and growth SFs, respectively \cite{Sebastian}. However, the diffraction profile of a  reciprocal space $q$-scan is also affected by the broadening effect caused by mosaicity; the observed broadening must therefore be deconvolved during this analysis. A summary of the possible broadening effects for different peaks is shown in Table~\ref{tab:QscanTable}.

\begin{figure}[!htbp]
\includegraphics[width=0.48\textwidth]{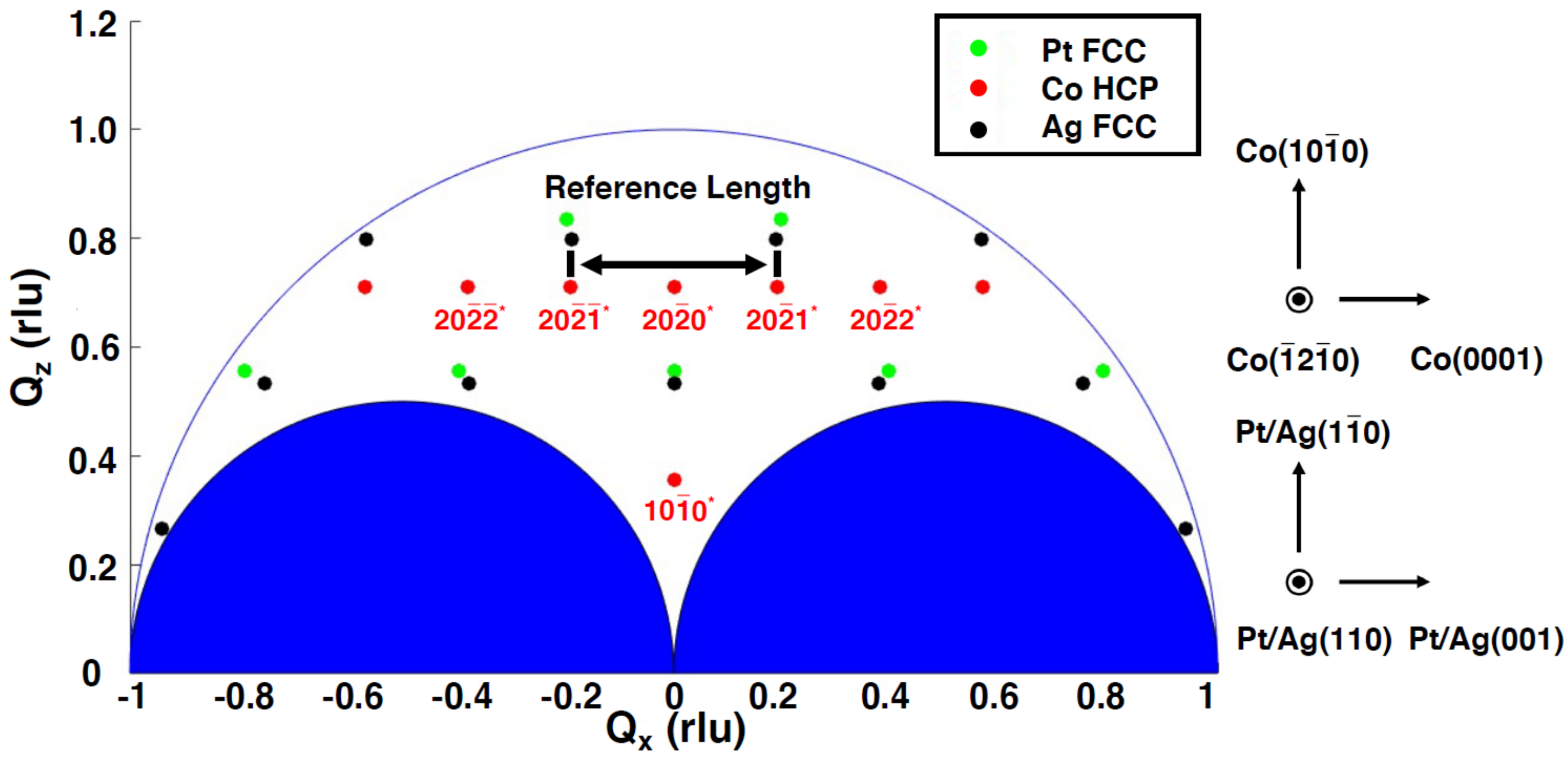}
\caption{\label{tab:reciprocalSpaceMap}
Reciprocal space map for the deposited thin film. The orientation of the film is indicated by the coordinate axes. The reciprocal space distance between  Co\hkl(20-2-1) and \hkl(20-21) peaks is marked as the Reference Length which is used for peak width normalization for $w_{h}$ values shown in Table~\ref{tab:QscanTable}. The limitations of the instrument prohibit the probing of regions within the shaded blue areas and beyond the blue arc.}
\end{figure}

\begin{figure}[!htbp]
\includegraphics[width=0.5\textwidth]{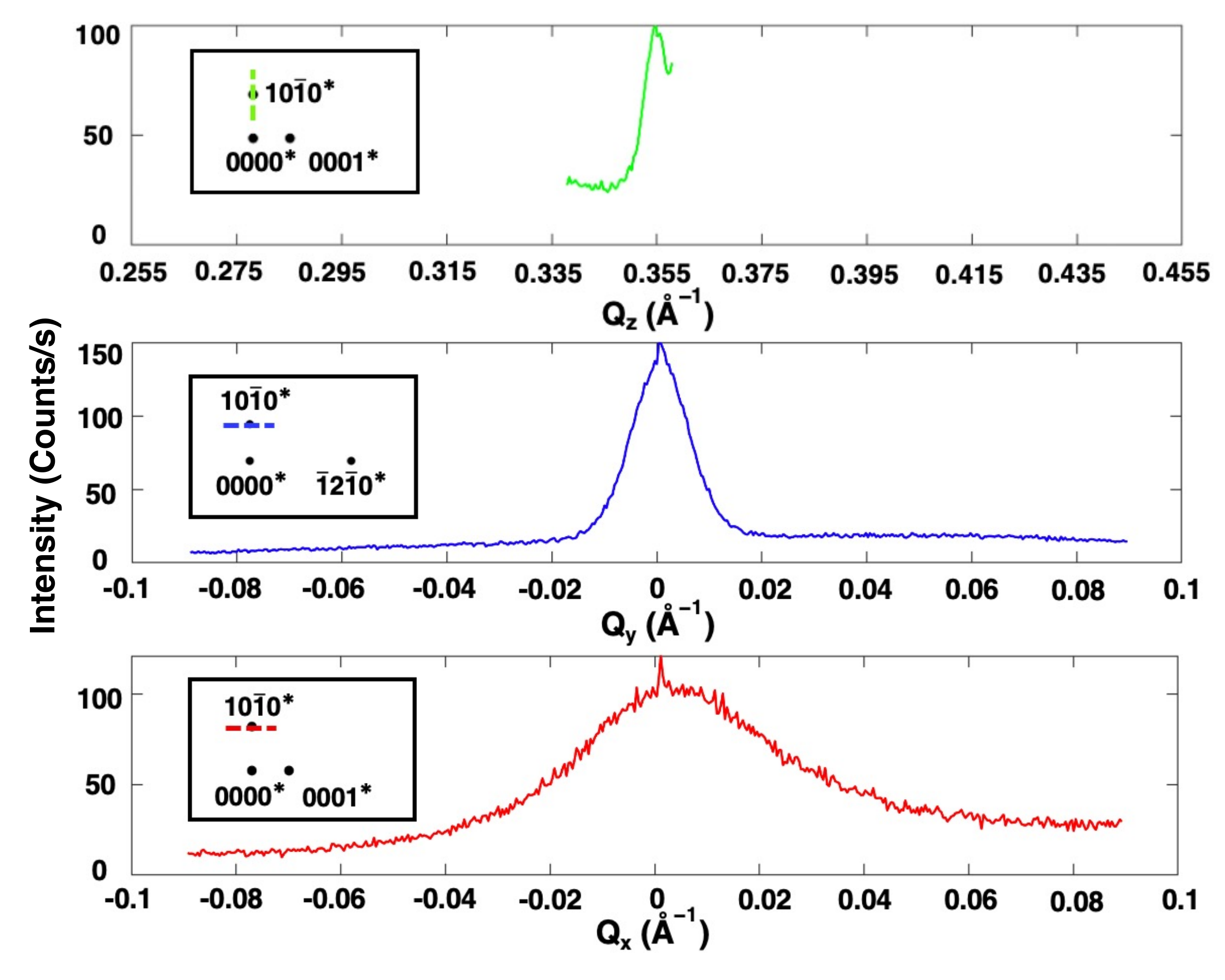}
\caption{\label{tab:Qscan}
Diffraction profile of reciprocal space $q$-scans performed on the Co\hkl(10-10) peak of Co(20 $nm$)/Pt thin film along the  Co\hkl(10-10), Co\hkl(-12-10), and Co\hkl(0001) plane normal directions. The inset of each figure is the reciprocal space map which shows the identity of the measured peaks and the dashed line indicates the scanning direction of X-ray. For clarity, the rightmost section of the scan along the Co\hkl(10-10) normal direction was not shown, as it is obscured by the Si\hkl(220) substrate peak (shown in Figure 1).}
\end{figure}

To characterize the broadening effects of SFs and mosaicity, X-ray reciprocal space $q$-scans were performed on the Co\hkl(20-20), \hkl(10-10), and \hkl(20-21) diffraction peaks along multiple directions for each sample. As a reference, the reciprocal space map of thin films is shown in Figure~\ref{tab:reciprocalSpaceMap}. Figure~\ref{tab:Qscan} presents measured reciprocal space q-scan diffraction profiles, where the dash lines in the insets indicate the scanning directions. The $w_h$ of each peak was estimated by fitting peaks with Gaussian functions. The results were then normalized with respect to the experimentally measured spacing between Co\hkl(20-2-1) and \hkl(20-21) peaks in reciprocal space; this spacing is marked in Figure~\ref{tab:reciprocalSpaceMap} as the ``Reference Length". Table~\ref{tab:QscanTable} shows the normalized $w_h$ values. The normalized $w_h$ of the  cobalt \hkl(10-10) peak along the \hkl(10-10) plane normal was used as a baseline for the other peak widths, as it is not affected by the broadening effects from either SFs or mosaicity. The broadening effects caused by SFs can be determined by comparing the diffraction profile of the Co\hkl(10-10) diffraction peak along the Co\hkl(10-10), \hkl(-12-10), and \hkl(0001) plane normals. As shown in Figure~\ref{tab:Qscan} and Table~\ref{tab:QscanTable}, the $w_h$ values of the Co\hkl(10-10) and \hkl(20-10) diffraction peaks along Co\hkl(0001) plane normal are considerably larger than those measured along the \hkl(-12-10) and \hkl(10-10) plane normals, indicating the presence of stacking faults.

\begin{table}[!htbp]
\caption{\label{tab:QscanTable}
Broadening effects present and $w_h$ values for cobalt peaks in the Co (20 $nm$)/Pt and Co (50 $nm$)/Pt thin films measured by reciprocal space $q$-scans along primary directions. S denotes the broadening effect caused by stacking faults; M refers to the broadening effect caused by mosaicity; and N signifies the absence of both effects. The values were normalized with respect to the reciprocal space distance between Co\hkl(20-2-1) and Co\hkl(20-21) peaks.}
\begin{ruledtabular}
\begin{tabular}{cccccc}
 & Co & Scan & Broadening & $w_h$ & $w_h$\\
 & Peak & Direction & Effects & (20 nm) & (50 nm)\\
\hline
     & \hkl(10-10) & \hkl(0001) & S+M & $0.1071$ & $0.1360$\\
      & \hkl(10-10) & \hkl(-12-10) & M & $0.0341$ & $0.0487$\\
      & \hkl(10-10) & \hkl(10-10) & N & $0.0149$ & $0.0099$\\
      & & & & &\\
     & \hkl(20-20) & \hkl(0001) & S+M & $0.1553$ & $0.0683$\\
      & \hkl(20-20) & \hkl(-12-10) & M & $0.0550$ & $0.0578$\\
      & \hkl(20-20) & \hkl(10-10) & N & $0.0188$ & $0.0177$\\
      & \hkl(20-21) & \hkl(0001) & S+M & $0.0671$ & $0.0675$\\
\end{tabular}
\end{ruledtabular}
\end{table}

For each sample, ${\Delta}w_{h, sf}$ was calculated as the difference in normalized $w_h$ between Co\hkl(10-10) peak widths measured along the \hkl(0001) and \hkl(-12-10) plane normals. By solving Equation \eqref{eqn:equation 1a}, the sum of deformation and growth SF densities in each sample was calculated, as shown in Table~\ref{tab:StackingFaultDensities}. Our calculations reveal considerable SF densities of 7.6\% and 9.1\% for our 20 nm and 50 nm films respectively, suggesting that SF density increases with thickness within this thickness regime. In principle, the individual densities of deformation and growth SFs can be calculated by comparing the $w_{h, sf}$ values for the \hkl(20-20) and \hkl(20-21) peaks. (The \hkl(10-11) peaks are inaccessible, as indicated in Figure 4.) However, measuring the \hkl(20-21) peak width along the \hkl(-12-10) normal direction requires a change in film orientation that would introduce additional broadening, thereby compromising our analysis. Moreover, the unusual thickness trends for the \hkl(20-20) peak widths warrant further investigation.

\begin{figure}[htbp]
\includegraphics[width=0.45\textwidth]{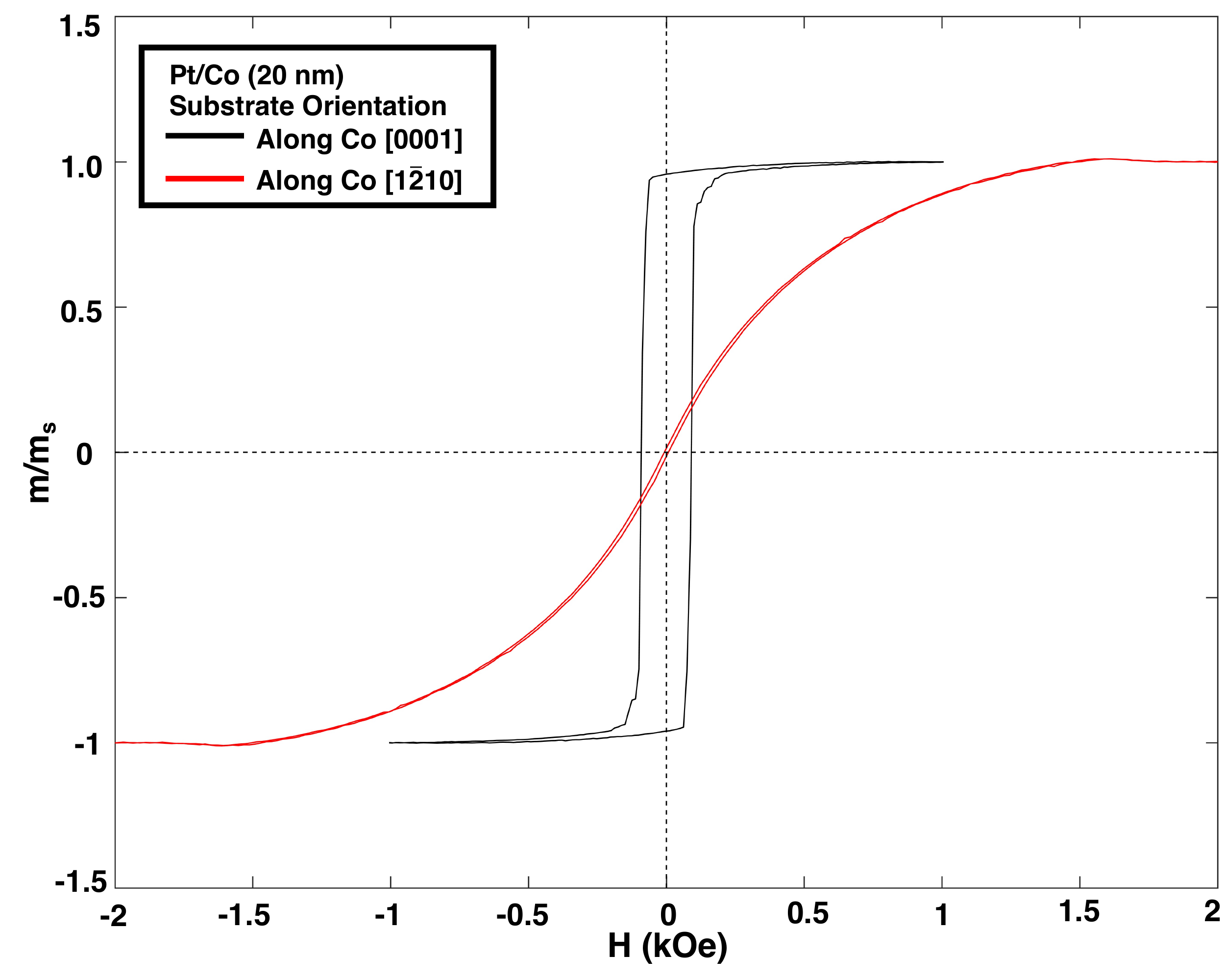} \\
\includegraphics[width=0.45\textwidth]{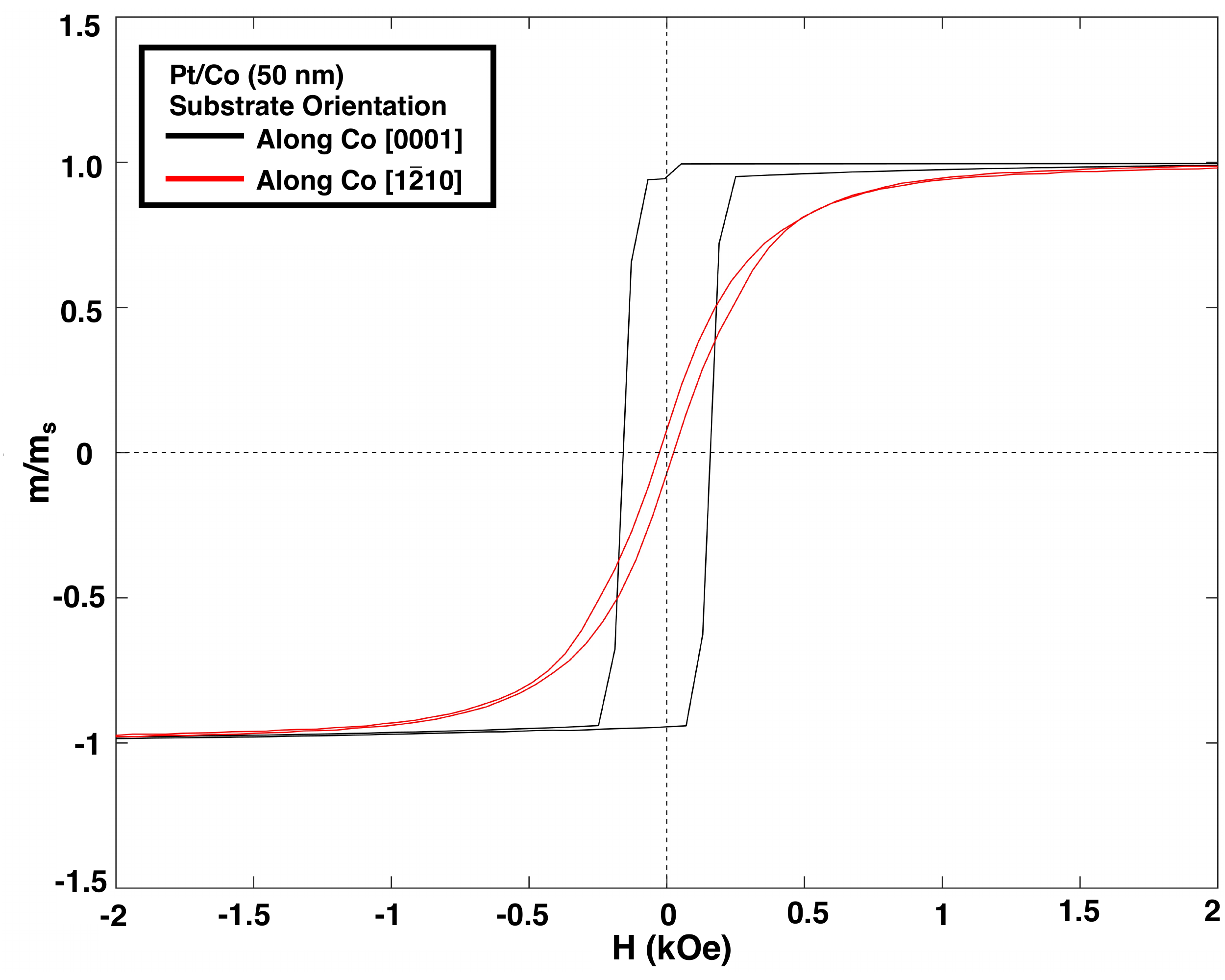}
\caption{\label{tab:Agm20nm}
Measured $M$\textendash $H$ hysteresis curves for Co/Pt films. The thickness of Co in each film in given in parentheses. The measurement directions are indicated with respect to Co.}
\end{figure}

Based on the characterized $M$\textendash $H$ hysteresis curves, the deposited thin films exhibit a strong in-plane anisotropy, as shown in Figure~\ref{tab:Agm20nm}. Seeing as the uniaxial symmetry axis (the c-axis, i.e., \hkl<0001>) of HCP Co lies within the film plane for Co\hkl(10-10) films, these results suggest that the MCA is dominant over any interface-induced anisotropy present. The anisotropy energy density, $K_u$, is shown in Table~\ref{tab:StackingFaultDensities} for the two films. In both cases, $K_u$ is considerably lower than the bulk $K_1$ value of 450 $kJ/m^{3}$ \cite{Cullity}, supporting the hypothesis that SFs lower the MCA energy of HCP Co. Moreover—as expected—$K_u$ decreases with increasing intrinsic SF density. Finally, we note that lattice strain along the Co\hkl[11-20] direction may also influence the values of $K_u$ magnetoelastically. 

Asymmetric ultrathin films based on all-FCC \hkl(111) Co/Pt layers possess a strong interfacial DMI stemming primarily from Co/Pt interface, as well as a strong perpendicular magnetic anisotropy (PMA) which is enhanced in the presence of Co/Ni layers. Beyoxpecting similar properties in ultrathin asymmetric Co\hkl(10-10)/Pt\hkl(110)–based films, we anticipate the emergence of DMI anisotropy, as well as the additional presence of a significant uniaxial in-plane MCA that is modulated by the degree to which the HCP phase of Co is preserved. By capitalizing on the tunability of a multilayer $C_{2v}$ Co/Pt system, one could explore the interplay between all three effects in FM/HM films\textemdash and potentially stabilize elusive chiral structures such as antiskyrmions and bimerons, the latter being the in-plane equivalents of skyrmions \cite{Gobel2019}.

\begin{table}[!htbp]
\caption{\label{tab:StackingFaultDensities}
The combined deformation ($\alpha$) \& growth ($\beta$) stacking fault densities for Co and the magnetocrystalline anisotropy energy density ($K_u$) in Co/Pt films differing in the thickness of Co ($t_{co}$).}
\begin{ruledtabular}
\begin{tabular}{cccc}
$t_{Co}$ & $\alpha$ + $\beta$ (\%) & $K_u(kJ/m^{3})$\\
\hline
20 nm & 7.644 & 64.8 \\
50 nm & 9.142 & 52.4 \\

\end{tabular}
\end{ruledtabular}
\end{table}

\section{\label{sec:conc}Conclusion}
In this work, we characterized Co(20nm)/Pt and Co(50nm)/Pt $C_{2v}$ films sputtered at room temperature on Si\hkl(110) substrates with a Ag growth seed. The epitaxial relationship between the Pt and Co layers was identified as HCP Co\hkl(10-10)\hkl[0001] $\parallel$ FCC Pt\hkl(110)\hkl[001]; to our knowledge, this is the first report of HCP $\parallel$ FCC epitaxy in an FM/HM film system. These films exhibit a definitive uniaxial in-plane magnetic anisotropy along Co\hkl [0001], in agreement with the observed HCP \hkl(10-10) texture of Co. An analysis of the broadening of Co\hkl(10-10) peaks along principal directions revealed stacking fault densities of 7.6\% and 9.1\% in the 20 nm and 50 nm films, respectively. Both films have significantly lower $K_u$ values compared to bulk cobalt, which we attribute\textemdash at least in part\textemdash to the appreciable intrinsic stacking fault densities observed. We anticipate that the observed epitaxial relationship and the singular in-plane easy axis will be preserved in thinner films\textemdash where the collective impact of interface-induced PMA, anisotropic DMI and in-plane uniaxial anisotropy on the stability of various chiral spin structures in multilayer FM/HM films could be explored. 

\begin{acknowledgments}

All authors acknowledge the support of the Defense Advanced Research Agency Program (DARPA) on Topological Excitations in Electronics (Grant No. D18AP00011) and the use of the Materials Characterization Facility at CMU, supported by grant MCF-677785. M.D.K. is also grateful for the financial support of the National GEM Consortium as well as the Neil \& Jo Bushnell and Bradford \& Diane Smith Engineering Fellowships, all awarded by the College of Engineering at CMU.
\end{acknowledgments}

%

\nocite{*}
\bibliography{Main.bib}

\end{document}